\documentclass[usenatbib]{mnras}

\usepackage[T1]{fontenc}
\usepackage{aecompl}
\usepackage{graphicx}
\usepackage{amsfonts}
\usepackage{amsmath}
\usepackage{rotating} 
\usepackage[varg]{txfonts}
\usepackage{fixltx2e} 
\usepackage{url}
\usepackage{etoolbox}
\usepackage{color}
\usepackage{hyperref}
 \hypersetup{
    colorlinks,%
    citecolor=magenta,%
    filecolor=blue,%
    linkcolor=magenta,%
    urlcolor=blue
}
\providecommand{\adsurl}[1]{}

\providecommand\aap{A\&A}            
\providecommand\mnras{MNRAS}

\providecommand\prd{PRD}

\providecommand\jcap{JCAP}

\providecommand\grg{Gen.~Rel.~Grav.}

\providecommand\annrevnucpartphys{Ann.~Rev.~Nucl.~Part.~Sci.}

\providecommand\cqg{Class.~Quant~Gravity}   %

\begin{document}

\title[Backreaction in Newtonian cosmology]{On Backreaction in Newtonian cosmology}

\author[T. Buchert]{Thomas Buchert\\
Univ Lyon, Ens de Lyon, Univ Lyon1, CNRS, Centre de Recherche Astrophysique de
  Lyon UMR5574, F--69007, Lyon, France}

\maketitle

\begin{abstract}
We clarify that a result recently stated by Kaiser 
is contained in a theorem of Buchert and Ehlers that is widely known for its main result: 
that there is no global kinematical backreaction in Newtonian cosmology. 
Kaiser cites this paper, re-derives parts of the theorem, but incompletely restates its content. He makes
further claims, which cannot be proven beyond the limited
context of Newtonian cosmology. 
We also discuss recent papers of R\'acz et al. and Roukema who claim
the existence of global backreaction within the Newtonian framework.
\end{abstract}

\begin{keywords}
cosmological parameters -- cosmology:theory -- dark energy
\end{keywords}

\section{Average properties of Newtonian models} \label{s-intro}

In a recent paper \citeauthor{Kaiser17b} (\citeyear{Kaiser17b}; hereafter K17)
considers, as do \citeauthor{BuchertEhlers97}
(\citeyear{BuchertEhlers97}; hereafter BE---see also \citealt{EhlersBuchert97}) the Euler--Poisson system in the fluid
approximation for a ``dust'' matter model in the mean field
approximation of Newtonian gravity, Eqs.\ (2) and (3) in K17 ({\it cf.} point [B] below).
Buchert and Ehlers performed spatial averaging of the kinematical
scalars of the system to obtain the following general expansion law,
given in BE, Eq. (B4):
\begin{eqnarray}
\label{general}
3{{\ddot a}_{\cal D} \over a_{\cal D}} + 4\pi G \frac{M_{\cal D}}{a_{\cal D}^3 } -
\Lambda =  :{\cal Q}_{\cal D} = \qquad\qquad\qquad\nonumber\\ 
\langle\boldsymbol{\nabla}\cdot \lbrack {\mathbf u} (\boldsymbol{\nabla} \cdot {\mathbf u} ) -
({\mathbf u} \cdot \boldsymbol{\nabla}) {\mathbf u} \rbrack \rangle_{\cal D}
- \frac{2}{3} \langle\boldsymbol{\nabla} \cdot {\mathbf u} \rangle_{\cal D}^2 \nonumber\\
+2\left(\Omega^2 - \Sigma^2 \right) +
2\left(\Omega_{ij}\langle\hat{\omega}_{ij}\rangle_{\cal D}
- \Sigma_{ij}\langle\hat{\sigma}_{ij}\rangle_{\cal D}\right),
\end{eqnarray}
where $a_{\cal D} \propto V_{\cal D}^{1/3}$ denotes the volume scale factor of {\it any} compact averaging domain $\cal D$ with volume $V_{\cal D}$, $\langle \cdot \rangle_{\cal D}: = 1/ V_{\cal D} \int_{\cal D}{\mathrm d}V\cdot$ the averaging operator, $\mathbf u$ the peculiar-velocity field (with its shear $\hat{\sigma}_{ij}$ and vorticity $\hat{\omega}_{ij}$), defined with respect to a global reference background flow, usually taken to be a Hubble flow (and Kaiser recalls correctly that this reference flow is {\it a priori} not fixed to be a Hubble flow).
Omitting global shear $\Sigma$ and global vorticity $\Omega$ of the model universe,
we arrive at the result of Kaiser (i.e., the kinematical backreaction term, Eq. (9), in K17).\footnote{\label{f1}The three corrections [1a], [1b] and [1c] in \citep{Buchert_comment_v1} of Kaiser's re-derivation of part of BE's results in \citep{Kaiser17a} have now been included (apart from the erroneous factor $3/2$ in K17, Eq.~(9), which should read $2/3$) after consulting \citep{Buchert_comment_v1} (N. Kaiser, {\it priv. comm.}), except that Kaiser still claims that his derivation is more general.}

Kaiser (K17; App.A) re-derives the result of BE stating that ``we show how this may be obtained directly''.
Inspection of his derivation shows, however, that it is essentially identical to the line of reasoning of BE, but less straightforward.
Kaiser obtains a linear, background-dependent term that (as he now appreciates, see footnote~\ref{f1}) is zero. The direct derivation of BE does not lead to such a zero term.
Follow-up papers and reviews also provided complementary derivations and
thorough discussions of the properties of the kinematical backreaction term [e.g., \cite{Buch00}; \cite{Buchert08status}, Sect.~3; \cite{BuchRas12}, Sect.~2.5]. 

The result (\ref{general}) is valid for all compact domains from an infinitesimal domain up to any scale, e.g., up to the boundary of a compact model universe, and the second term in the second line of (\ref{general}) is there on all scales.
Kaiser's paper titled {\it ``Why there is no Newtonian backreaction''} is an improvement over the title {\it ``There is no kinematic backreaction''} of \citep{Kaiser17a}, but still overstates and misleadingly interprets the content of the BE theorem for the following reasons:
\begin{itemize}
\item[A)] What BE showed, and what Kaiser re-derives, is that there is no
  {\it global} kinematical backreaction in {\it Newtonian cosmology}
  for boundary-free compact model universes.\footnote{We shall explain this in point [C]. 
  Note that kinematical backreaction also
  vanishes for spherical regions embedded into a reference background
  \citealt{BKS00}, \citealt{Buchert11Towards}: Sect.~7.2. This latter
  provides a compact proof of the Newton iron sphere
  theorem. Kaiser quotes instead the work of \citet{EinsteinStraus} which is out of context. That backreaction vanishes for a {\it single} spherical region is well-known.} There is, however, kinematical backreaction, describing
  cosmic variance of the deviations from the assumed reference
  background, as Kaiser correctly states, and which is explored
  in the paper by \citealt{BKS00}. Thus, the suggestion that there is ``no Newtonian
backreaction'' {\it per se} is not supported by the text.
  Kinematical backreaction is present in Newtonian models
  in the {\it interior} of, e.g., a 3--torus model\footnote{Kaiser confuses the scale-dependent volume expansion rate $H_{\cal D}$ with the background expansion rate $H$. It is trivial that 
``There is no Newtonian backreaction on $a(t)$ from structure'' (K17), but there is Newtonian backreaction on $a_{\cal D}(t)$ from structure, except on the global scale where $a_{\cal D}(t)$ reduces to $a(t)$.
Kaiser discusses, how the equation for $a(t)$ can be chosen to recover the usual Newtonian equations
for point particles, but this does not imply that the average motion of the point particles follows the same $a(t)$. This dismisses the non-local aspect of averaging contained in $a_{\cal D}(t)$.} (and the expectation values of the peculiar-velocity invariants are in general non-zero on regional domains). The term ${\cal Q}_{\cal D}$ in (\ref{general}) can be expressed
 in terms of Minkowski Functionals \citep[][Sect.~1.11]{Buch00}, \citep[][Sect.~3.1.2]{Buchert08status}. 
 These functionals describe morphological properties of the density distribution
 and depend on all correlation functions of a density field arising
 out of smoothing a point-like distribution. Consequently, even small-amplitude interior
 backreaction terms are important in characterizing the cosmic web.\\

\item[B)] Kaiser, in K17, Sect.~2, leaves the impression that a more elementary approach using discrete particles leads to the same result as that obtained from the Euler--Poisson system for a dust fluid in the mean field approximation. This impression comes from the comparison of the Newtonian equation of motion for particles with the Euler equation (while he re-derives the main result, K17, Eq.~(9), within a fluid mechanical approach, K17, Sect.~3 and App.A). Such an approach has been discussed by \citet{GibbEllis13Newt}, who show\footnote{Kaiser now cites this paper without pointing out the limitations explained 
by \citet{GibbEllis13Newt}.} that a set of {\it central configuration constraints} has to be also satisfied. It should be clear that the {\it general} link between a particle\- approach and the governing fluid equations is more subtle: starting from discrete particles, coarse-graining the Klimontovich density in phase space and taking velocity moments leads to extra anisotropic multi-stream stresses modifying the Euler equation, and deviations from mean field gravity modifying the Poisson equation, see, e.g., \citep{Buchert05adhesive}, not done in K17 and BE. This also leads to extra backreaction terms.\footnote{In general relativity, the averaged equations have been given by including isotropic stresses \citep{Buch01scalav}.} There is no ``significant advance over the approach followed e.g. in BE'' (K17):
purely space-dependent arguments already assume the vanishing of the above terms.\footnote{Recall that in situations of multiple streams, space-dependent variables cease to be functions and we have to move to a description in phase space.} Only including these terms would render the result more general.\footnote{\label{foot7}Keeping the mean field assumption it is straightforward to include velocity dispersion: taking velocity moments of the phase space density, the Euler-Jeans equation replaces the Euler equation featuring multi-stream stresses $\Pi_{ij}$. The additional source, $-\frac{1}{\varrho}\frac{\partial}{\partial x_j}\Pi_{ij} =: - \Psi_i$, results in a divergence term, $- \langle \partial \Psi_k / \partial x_k \rangle_{\cal D}$, adding to the kinematical backreaction term in (\ref{general}), and leaving the main result of BE on the global vanishing of the kinematical backreaction term unchanged. There is no such term in Kaiser's derivation.}\\

\item[C)] Kaiser omits a discussion of the {\it
  uniqueness} of solutions to the Poisson equation.  This problem,
  discussed in BE, is crucial to Newtonian cosmology since we have to
  specify boundary conditions. BE considered various possibilities,
  including cases where kinematical backreaction
  does not globally vanish (Charlier--type models),
  among them the possibility of introducing a background and
  deviations thereof defined on a 3--torus.
BE also discuss why this latter is the only relevant possibility for Newtonian cosmological models (apart from Charlier-type models without a homogeneity scale). To show this
  we recall that the Poisson equation does not uniquely define the
  potential. It is unique if we require square integrability of the
  potential in the case of $\mathbb{R}^3$. (The
    sum in K17, Eq.~(5) is in general divergent, a problem of which
    Newton and Einstein were well aware
    \citep[see, e.g.][Sect.~1.1, and references therein]{GibbEllis13Newt}.)
  The possibility
    of a square integrable potential over $\mathbb{R}^3$ is, however, irrelevant for
  cosmology.
The potential is also
  unique if the spatial average of the source vanishes. Introducing a
  background and deviations thereof allows one to consider periodic
  peculiar-velocity fields and density contrasts with a
  vanishing average of the source of Poisson's equation for a
 periodic peculiar-potential (a 3--torus model). This
  architecture, being identical to that of Newtonian
  $N$-body simulations, implies that the peculiar-potential
    exists and is unique apart from the addition
    of a solution of the Laplace equation, whose only periodic solutions are spatially constant.
  This constant term can be removed by the translation
 invariance of Newton's equations and set to zero without loss of
  generality.

In general relativity (GR)
these problems do not arise and a model universe
(that is, in general, background-free) is not restricted to
obey this global constraint.  In addition, the kinematical
backreaction variable couples to the averaged intrinsic
curvature [Sect.~3.2.2 ff in \citep{Buchert08status}], removing the ``Newtonian
anchor'' [Sect.~3.2.3 in \citep{Buchert08status}, Sect.~2.5 in \citep{Buch00}], which enforces the vanishing of structure averages globally on an assumed
background. This is the reason why the globally vanishing backreaction in Newtonian cosmology is ``by construction'': Kaiser misinterprets this insightful remark and notes that ``... ${\cal Q}_{\cal D}$ tends to zero very rapidly in the limit of large volumes regardless of whether the structure is assumed to be periodic." As we pointed out above, periodicity is a necessary element of the architecture of Newtonian models applied to cosmology;  by the very definition of ${\cal Q}_{\cal D}$, the term decays rapidly with volume, but it is crucial that it has to vanish exactly on the periodicity scale. We note the importance of a non-vanishing but small
${\cal Q}_{\cal D}$ for the evolution of averaged scalar curvature in the GR context, see the recent paper by \citet{Bolejko17a} who illustrates this by employing exact GR solutions.
Scale-dependent deviations from an assumed background are unavoidable.\footnote{\label{foot8}In GR this holds globally and can be traced back to first principles like the non-conservation of
intrinsic curvature [Eq. (13) in \citep{Buch00scalav}, \citep{Buchert08status}, and \citep{buchertcarfora}].}
\end{itemize}

\section{Concluding Remarks and Discussion}

We conclude that there is no new result in \citet{Kaiser17b}. 
Known for twenty years, the result of \citet{BuchertEhlers97} is key to Newtonian cosmology, and is uncontroversial among researchers working
on backreaction problems.

However, Kaiser's re-statement that Newtonian models cannot lead to non-vanishing backreaction is correct {\it globally}, and relevant for the rebuttal of recent work by \cite{nbody} on backreaction calculated within the Newtonian framework. We therefore discuss this latter work, as well as a recent complementary work by \cite{boud17}, and we point out why the conclusions drawn from these models on the possibility of global backreaction contradicts the result of \citet{BuchertEhlers97}.

Since kinematical backreaction is non-vanishing {\it in the interior of} a Newtonian simulation, 
and attains substantial values by going to smaller but still cosmologically relevant scales \citep{BKS00},
it is unsurprising that \cite{nbody} re-discover expansion variance by implementing a multi-scale volume partitioning in N-body simulations.
These toy-simulations may be interpreted to illustrate the effect on differential expansion properties, and only on a restricted range of scales. It is clear, however, and a consequence of the theorem (\ref{general}), that there {\it cannot be} backreaction in these simulations {\it globally}. 

The key-issue to understand this fact is the non-local nature of the gravitational interaction. \cite{nbody} adopt {\it assumptions} on individual domains of the global model universe. In their paper the assumption of vanishing backreaction within a spherical idealization of domains is adopted. The non-locality of gravitation that is reflected by global constraints (boundary conditions in Newtonian cosmology and propagating spatial constraints in GR 3+1--foliated models) assures, however, that any property of the regional domain is ``known'' to the rest of the model universe. Thus, any approximate assumption made must be rendered consistent with the whole model universe. Otherwise, an error is introduced that---as in the case of \cite{nbody}---results in a fictitious global contribution of backreaction. To avoid this error,
the generic situation has to be implemented, accounting in particular for backreaction on regional domains (i.e. deviations from spherical symmetry). It is of equal
importance that the domains of the volume partition are be joined properly to obtain consistency with
the global constraint. Furthermore, although
an evolution of the background away from a homogeneous Hubble flow is a generic property of GR cosmologies on any scale,\footnote{The quantitative importance of the backreaction effect in GR is a result of the observation that the average model evolves away from a pre-defined FLRW (Friedmann-Lema\^\i tre-Robertson-Walker) background due to the non-conservation of curvature (see footnote~\ref{foot8}). 
A scale-dependent ``background'' (affected by structure formation) has been investigated in the GR context \citep{RB}, together with global instability properties of the FLRW model \citep{RBCO}.} it lacks a physical basis within the Newtonian framework.

The {\it separate universe conjecture} (\cite{separateuniverse} and references therein), i.e. making restrictive assumptions on regional domains and considering the region to evolve separated from the environment, has to be put under careful scrutiny, especially in nonlinear modelling attempts. 

Another example for an uncontrolled implementation of this conjecture is the recent paper by \cite{boud17} 
that suggests an alternative explanation for
structure-emerging global backreaction within the Newtonian framework.
While \cite{boud17} uses generic properties on subdomains, in particular including backreaction on regional domains,\footnote{\cite{boud17} employs the analytical backreaction model of \cite{BKS00} that is based on exact average properties, and an analytical model for the evolution of inhomogeneities that has earlier been well-tested against N-body simulations.} he implements the assumption of ``silent virialization''\footnote{The word ``silent'' here refers to the separate universe conjecture rather than to so-called ``silent model universes'' in GR, ({\it cf.} \cite{Bolejko17b} and references therein), that have vanishing magnetic part of the Weyl tensor. Note that even ``silent model universes'' do not obey the separate universe conjecture, since the spatial GR constraints and their consistent propagation have to be obeyed which again reflects the non-locality of the gravitational interaction.} by stabilizing collapsed domains with the consequence of obtaining an excess global expansion over the assumed background expansion. However, virialization in collisionless Newtonian systems can be modelled exactly by including velocity dispersion, as is explained in point (B) of section~\ref{s-intro}. This {\it cannot} lead to a global backreaction, as is explained in footnote~\ref{foot7}. 

Furthermore, it does not help to appeal to the algebraic similarity between the Newtonian backreaction model of \cite{BKS00} and the corresponding relativistic model of \cite{rza2} through the claim that ``virialization" induces an excess of expansion in a ``relativistically realistic'' \citep{boud17} situation, while
the actual realization of the model is still confined to the Newtonian setting. 

Also in relativistic N-body simulations (that include virialization), the spatial GR constraints and their propagation has to be controlled, representing what in Newtonian theory are the boundary conditions. These simulations may also face the constraints discussed here for Newtonian models, if periodic boundary conditions on deviations from a fixed background model are implemented. 

A promising route to background-free investigations of the average properties of GR universe models is furnished by multi-scale volume partitions that
have been investigated during the last decade (a comprehensive list of works may be found in recent reviews, e.g., in \citealt{BuchRas12}, \citealt{W14}). In particular, the exact volume partitioning formulas for GR models given in \cite{buchertcarfora} and \cite{multiscale} can be profitably used to demonstrate the consequences of any regional assumption made. (The present debate motivated us to explicitly demonstrate these consequences for an exact volume partition of Newtonian models in a forthcoming paper.)

It may well be possible to model regional domains of the real Universe using Newtonian equations, but this does not provide justification to model the entire observable Universe by a single Newtonian solution. Kaiser's remarks appear to assume that small regions combine into a single cosmology in a trivial way (i.e. that the ``background'' of all small regions should be identical). A kaleidoscopic modelling of the Universe by consecutive Newtonian domains restricts the generality substantially, for example, by calculating the average Ricci curvature of the model universe.

Kaiser's discussion of the relevance of the BE theorem for GR constitutes an opinion.
More efforts to justify that the Universe is everywhere close to the same background model have been invested by \citet{GW}. 
However, their assumptions are too restrictive
to apply to cosmological backreaction: see \citep{GWdebunk15} and
references therein, which provide an extensive discussion of the physics
of cosmological backreaction.

Finally, quoting statements of early papers that pioneered the subject
risks being anachronistic if developments in the literature over the subsequent
fifteen years are not taken into account. (The remarks cited by Kaiser are still
agreed upon today, but nuanced.)

\section*{Acknowledgments}
This work is part of a project that, in the final stage, has received funding from the European Research Council (ERC) under the European Union's Horizon 2020 research and innovation programme (grant agreement ERC advanced grant 740021--ARTHUS, PI: T. Buchert).
Thanks to J\"urgen Ehlers in memoriam with whom I share the results of BE, and thanks to Nick Kaiser, Martin Kerscher, Pierre Mourier, Jan Ostrowski, Boud Roukema, Syksy R\"as\"anen and David Wiltshire for correspondence, discussions and feedback, as well as to an anonymous referee for an insightful and very careful report.

\end{document}